\documentclass[12pt]{article}
\usepackage[a4paper, hmargin=2.5cm, vmargin=2.5cm]{geometry}
\usepackage{amsfonts,amsmath}
\usepackage[authoryear]{natbib} \bibpunct{(}{)}{,}{a}{,}{,}
\usepackage{comment}
\usepackage{multirow}
\usepackage{graphicx}
\usepackage{algorithm, algorithmicx, algpseudocode}

\newcommand{\bY}{\mbox{\boldmath $Y$}}
\newcommand{\bZ}{\mbox{\boldmath $Z$}}
\newcommand{\be}{\mbox{\boldmath $e$}}
\newcommand{\bs}{\mbox{\boldmath $s$}}

\newcommand{\by}{\mbox{\boldmath $y$}}

\newcommand{\bPsi}{\mbox{\boldmath $\Psi$}}
\newcommand{\bdelta}{\mbox{\boldmath $\delta$}}
\newcommand{\bDelta}{\mbox{\boldmath $\Delta$}}

\newcommand{\bSigma}{\mbox{\boldmath $\Sigma$}}
\newcommand{\bOmega}{\mbox{\boldmath $\Omega$}}
\newcommand{\bLambda}{\mbox{\boldmath $\Lambda$}}

\newcommand{\bmu}{\mbox{\boldmath $\mu$}}
\newcommand{\btheta}{\mbox{\boldmath $\theta$}}
\newcommand{\bI}{\mbox{\boldmath $I$}}
\newcommand{\bzero}{\mbox{\boldmath $0$}}

\begin{document}

\title{Multi-Node EM Algorithm for Finite Mixture Models}

%\author{Geoffrey J. McLachlan$^{1, \star}$, Sharon X. Lee$^2$}
\author{Sharon X. Lee$^{1}$, Geoffrey J. McLachlan$^{2, \star}$,
Leemaqz, K.L.$^2$}
\date{}

\maketitle

\begin{flushleft}
$^1$School of Mathematical Sciences, University of Adelaide, 
Adelaide, South Australia, 5005, Australia.\\
$^2$Department of Mathematics, University of Queensland, 
Brisbane, Queensland, Australia, 4072, Australia.\\
$^\star$ E-mail: g.mclachlan@uq.edu.au
\end{flushleft}

\begin{abstract}
Finite mixture models are powerful tools for modelling 
and analyzing heterogeneous data. 
Parameter estimation is typically carried out 
using maximum likelihood estimation via 
the Expectation-Maximization (EM) algorithm.  
Recently, the adoption of flexible distributions 
as component densities has become increasingly popular. 
Often, the EM algorithm for these models involves 
complicated expressions that are time-consuming to evaluate numerically. 
In this paper, we describe a parallel implementation of 
the EM-algorithm suitable for both single-threaded and 
multi-threaded processors and for both single machine 
and multiple-node systems. Numerical experiments are 
performed to demonstrate the potential performance gain 
in different settings. Comparison is also made 
across two commonly used platforms - R and MATLAB. 
For illustration, a fairly general mixture model is used 
in the comparison.  
\end{abstract}

%---------------------------------------------------%

\section{Introduction}
\label{sec1}

In recent years there has been an increasing use of 
finite mixtures of flexible distributions for 
the modelling and analysis of heterogeneous 
and non-normal data \cite{mlr19}. 
These models adopt component densities 
that offer a high degree of flexibility 
in distributional shapes. In particular, 
the skew-symmetric family of distributions, 
which includes the classical skew normal 
and skew $t$-distributions, 
has become increasingly popular \cite{J402,J027,J028,J066,J116,J164}. 
It has enjoyed applications in a range of areas 
including astrophysics, bioinformatics, biology, 
climatology, medicine, finance, fisheries, 
and social sciences 
\cite{J118,as12,tccg19,J004,J356,hagct19,J119,J184,hhe18}.

Traditionally, finite mixture models are fitted by 
maximum likelihood estimation, carried out via 
the Expectation-Maximization (EM) algorithm \cite{J034}. 
For simple component densities like the normal and 
$t$-distributions, the E- and M-steps are usually 
quite straightforward. But for some flexible distributions 
such as the skew normal and skew-$t$ mixture models, 
the E-step often involves complicated expressions; 
see the aforementioned references. 
For example, the conditional expectations in the E-step 
may require calculation of the moments of truncated 
(multivariate) distributions. 
Depending on the particular characterization 
of the component densities, this may involve numerical evaluations of 
multidimensional integrals that are computationally demanding.       

To speed up the model fitting process, a number of recent works 
have presented modified versions of the EM algorithm for 
parallel computing. The vast majority of these contributions 
are aimed at large scale distributed and/or cloud platforms 
such as GraphLab, Piccolo, and Spark; 
see, for example \cite{J600,J601,J602,J603,J604}. 
Relatively few have focused on smaller scale environments 
with a single machine or a small local network of machines. 
The paper \cite{J452} presented a simple parallel version 
of the EM algorithm for single machine, 
taking advantage of multithreading. 
For a $g$-component mixture model, 
the authors proposed to split the computations across $g$ threads. 
We shall refer to this as the multi-EM algorithm. 
The advantage of this approach is its simplicity and 
ease of implementation, as it requires minimal modification 
to the original (serial) code. 
More recently, \cite{llm19} presented a block EM algorithm 
where the data are horizontally split into $b$ blocks, 
allowing for an arbitrary number of threads to be used. 
Their algorithm was illustrated on multi-core machines.   
Note that both the multi-EM and block-EM 
algorithms were implemented in R and aimed at single 
(standalone) machines.      

In this paper, we describe another parallel implementation 
of the EM algorithm that is suitable for both single 
and small networks of machines. The structure of the EM algorithm 
allows easy splitting of the E-step into a single thread 
for each single observation and component. 
The M-step can also be naturally split into $g$ separate threads. 
Depending on the physical system used, 
the user may choose an appropriate number of threads to use 
for the E- and M-steps.  
For illustration, we adopt the finite mixture of 
canonical fundamental skew $t$ (CFUST) distributions \cite{J008} 
to assess the performance of the parallel algorithms. 
In addition, we implemented the algorithms in 
two commonly used mathematical platforms, namely R and MATLAB, 
and compared the performance gain across these platforms.

%---------------------------------------------------%
\section{The EM algorithm for finite mixture models}\label{sec2}

Finite mixture models provide a convenient mathematical 
representation of heterogeneous clusters within the data. 
Formally, the density of a finite mixture model is 
a convex combination of component densities. 
Let $\bY$ denotes a $p$-dimensional random vector 
consisting of $p$ feature variables of interest, 
and $\by$ be a realization of $\bY$. 
Then the density of a $g$-component mixture model takes the form 
\begin{eqnarray}
f(\by; \bPsi) &=& \sum_{i=1}^g \pi_i f_i(\by; \btheta_i), 
\label{eq:FM}
\end{eqnarray}
where $\pi_i$ denotes the mixing proportion for component $i$, 
$f_i(\cdot)$ denotes the density of the $i$th component 
of the mixture model, and $\btheta_i$ denotes the vector of 
unknown parameters of the $i$th component, for $i=1, 2, \ldots, g$. 
The vector $\bPsi$ contains all the unknown parameters 
of the mixture model, that is, 
$\bPsi=(\pi_1, \ldots, \pi_{g-1}, \btheta_1^\top, \ldots, \btheta_g^\top)$. 
Note that the mixing proportions satisfy $\pi_i>0$ 
and $\sum_{i=1}^g \pi_i=1$.

Traditionally, the component density is taken to be 
the (multivariate) normal distribution, that is, 
$f_i(\cdot) = \phi_p(\by; \bmu, \bSigma)$ 
where $\bmu$ is a $p$-dimensional vector of location parameters 
and $\bSigma$ is a positive-definite scale matrix. 
For illustration purposes, we consider $f_i(\cdot)$ to be 
the CFUST density in this paper, which is given by
\begin{eqnarray}
f_{\mbox{\tiny{CFUST}}}(\by; \bmu, \bSigma, \bDelta, \nu)
	= 2^q \, t_p(\by; \bmu, \bOmega, \nu) \, 
		T_q \left(\bdelta^\top\bOmega^{-1}(\by-\bmu) 
		\sqrt{\frac{\nu+p}{\nu+d(\by)}}; \bzero, \bLambda, \nu+p\right), 
\label{eq:CFUST}
\end{eqnarray}
where $t_p(\cdot)$ denotes the $p$-dimensional $t$-density 
and $T_p(\cdot)$ denote its corresponding cumulative distribution function. In the above, 
we let $\bOmega = \bSigma+\bDelta\bDelta^\top$, 
$\bLambda=\bI_q-\bDelta^\top\bOmega^{-1}\bDelta$, 
and $d(\by)=(\by-\bmu)^\top\bOmega^{-1}(\by-\bmu)$ 
is the Mahalanobis distance between $\by$ and $\bmu$. 
We shall refer to finite mixtures of (\ref{eq:CFUST}) as FM-CFUST. 
The CFUST distribution is fairly flexible and contains, 
as special and/or limiting cases, many commonly used distributions 
including the normal, $t$, Cauchy, and skew normal distributions. 
These are obtained by letting $\bDelta=\bzero$ and $\nu\to\infty$; 
$\bDelta=\bzero$; $\bDelta=\bzero$ and $\nu=1$; and $\nu\to\infty$, 
respectively. In addition, several characterizations of 
skew normal and skew $t$-distributions are also 
special and/or limiting cases of (\ref{eq:CFUST}) \cite{J105}.

Estimation of $\bPsi$ for mixture models is typically undertaken 
by maximum likelihood via the EM algorithm. 
The EM algorithm begins with an initialization step, 
where an initial estimate $\bPsi^{(0)}$ of $\bPsi$ 
are computed from an initial partition or via other strategies. 
We then alternate the E- and M-steps until some stopping criterion is satisfied. 
To facilitate parameter estimation via the EM algorithm, 
a set of latent binary variables $\bZ_j$ is introduced, 
representing the component membership of $\by_j$ 
-- the $j$th observation in the data. 
More formally, $Z_{ij}=(Z_j)_i =1$ if $\by_j$ belongs to 
the $i$th component of the mixture model and zero otherwise. Depending on the choice of the component density, 
additional latent variables may be introduced 
to simplify calculations. In the case of FM-CFUST, 
these include latent gamma random variables 
and latent truncated normal random variables. 
The technical details of the EM algorithm for FM-CFUST 
are omitted here, but can be found in \cite{J164}. 
An outline of the procedure of the EM algorithm is given below. 
\begin{enumerate}
\item[1)] \emph{Initialization}: 
	Obtain $\bPsi^{(0)}$ from an initial partition 
	or some other starting strategies. 
	Calculate the initial log likelihood value 
	using the following with $k=0$:
\begin{equation}
\ell^{(k)} = \sum_{j=1}^n \log f(\by_j; \bPsi^{(k)}). 
\label{eq:logL}
\end{equation}
%Set $k=1$.
\item \emph{E-step:} 
	Calculate the posterior probability of component membership:
\begin{equation}
\tau_{ij}^{(k)} = \frac{\pi_i^{(k)} f_i(\by_j; \btheta_i^{(k)})} 
	{f(\by_j; \bPsi^{(k)})} 
	= \frac{f_{ij}^{(k)}}{f_j^{(k)}},
	\label{eq:TAU}
\end{equation}
for $i=1, 2, \ldots, g$ and $j=1, 2, \ldots, n$. 
Calculate any other required conditional expectations 
$\be_{1ij}^{(k)}$, $\be_{2ij}^{(k)}, \ldots$.  
\item \emph{M-step:} 
	Compute updated parameter estimates $\bPsi^{(k+1)}$ 
	based on the output of the E-step.
\item \emph{Stopping criterion:} 
	Update the log likelihood value using (\ref{eq:logL}). 
	Check whether the stopping criterion is satisfied. 
	If so, return the output of the M-step. 
	Otherwise, increment $k$ to $k+1$ and return to the E-step.  
\end{enumerate}

%---------------------------------------------------%
\section{A multi-node EM algorithm}

The computation of the conditional expectations in the E-step 
can be quite time consuming, especially for large values of $p$ 
(and $q$ in the case of FM-CFUST). 
We now describe a parallel implementation of the EM algorithm 
that is suitable for multi-core and/or multi-node operating systems. 
As can be observed from the structure of the EM algorithm 
outlined above, the computation of $\tau_{ij}^{(k)}$ 
and other conditional expectations in the E-step 
can be carried out separately for each $i$ and $j$. 
It is thus intuitive to split the data into $m$ blocks 
where $1\leq m \leq gn$. The value of $m$ can be user-specified 
or chosen to best match the physical systems. 
In R, $m$ needs to be specified explicitly, 
whereas for MATLAB $m$ will be set to the number of physical cores 
in the system by default. Note that the system may comprise of 
more than one machine and thus $m$ is the total number of 
physical cores across all machines.

%--------------------------------%
\subsection{Parallel initialization}

The EM algorithm is sensitive to starting values 
and thus it is important to choose good initial estimates 
for the parameters of the model. With $m$ threads, 
we can trial $m$ different initializations concurrently. 
The algorithm begins with $m$ different initial partitions 
that can be obtained, for example, by $k$-means or 
random partitions. Each of the $m$ threads computes $\bPsi^{(0)}$ 
based on the given initial partition. The initial log likelihood value 
$\ell^{(0)}$ is then computed. There is an option for each thread 
to compute a small number $r$ of (burn-in) EM iterations. 
However, for simplicity we have used $r=0$ in the numerical experiments, that is, without any burn-in iterations. 
A master thread then gathers the $\ell^{(0)}$ from the $m$ threads 
and selects the one with the smallest value of $\ell^{(0)}$ 
to provide the starting values for the multi-node EM algorithm.            

In the process of computing $\ell^{(k)}$, the quantities $f_{ij}^{(k)}$ 
and hence $f_j^{(k)}$ (\ref{eq:TAU}), were evaluated 
for $i=1, 2, \ldots, g$ and $j=1, 2, \ldots, n$. 
They correspond to the numerator and denominator of $\tau_{ij}^{(k)}$, 
respectively. Hence, the $\tau_{ij}^{(1)} = f_{ij}^{(0)} / f_j^{(0)}$ 
are also computed during the initialization step 
and are passed on to the E-step in the first iteration. 
A summary of the parallel initialization step 
is presented in Algorithm \ref{A1}.

%\begin{comment}

\begin{algorithm}
\caption{Parallel initialization step of multi-node EM algorithm}
\label{A1}
\begin{algorithmic}[1]
	\State Compute $m$ different initial partitions of the data.
  \For {thread $l \in \{1, 2, \ldots, m\}$} concurrently
		\State Compute $\bPsi^{(0)}$.
		\State Compute $f_{ij}^{(0)}$ for $i=1, 2\ldots, g$ and $j=1, 2, \ldots, n$.
		\State Calculate $f_j^{(0)} = \sum_{i=1}^n \pi_i^{(0)} f_{ij}^{(0)}$.
		\State Calculate $\ell^{(0)} = \sum_{j=1}^n \log f_j^{(0)}$. 
		\State Calculate $\tau_{ij}^{(1)} = f_{ij}^{(0)} / f_j^{(0)}$.
		\State (Optional) Run $r$ iterations of the EM algorithm. 
		\State Return $\ell^{(0)}$, $\bPsi^{(0)}$, and $\tau_{ij}^{(1)}$ to the master thread. 
  \EndFor
	\State Select $l$ that corresponds to the smallest $\ell^{(0)}$ from the results of the $m$ threads.
	\State Return the selected $\ell^{(0)}$, $\bPsi^{(0)}$, and $\tau_{ij}^{(1)}$.
\end{algorithmic}
\end{algorithm}

%\end{comment}

%--------------------------------%
\subsection{Parallel E-step}

Computation of the conditional expectations 
$\be_{1ij}^{(k)}, \be_{2ij}^{(k)}, \ldots$ can be performed 
in parallel for each $i$ and $j$. 
Depending on the total number of threads, 
if $m<ng$ then a thread $l$ may be responsible for 
more than one value of $i$ and $j$. 
For performance, preference would be given to $i$ 
so that $\tau_{ij}^{(k)}$ can be computed in the same thread 
for each $j$. Let $l = 1, 2, \ldots, m$ be 
the index of the threads and $J_l$ be the set of index 
(of observations) assigned to thread $l$. 
An outline of the E-step is presented in Algorithm \ref{A2}. 
During the E-step, each thread calculates 
$\tau_{ij}^{(k)}$, $\ell_l^{(k-1)}$, 
and $\be_{1ij}^{(k)}, \be_{2ij}^{(k)}, \ldots$ 
for observations $\by_j$ in $i \in J_l$.   
For the first iteration, $\tau_{ij}^{(1)}$ have been passed on 
from the initialization step and hence can be skipped. 
Unlike the traditional implementation of the EM-algorithm, 
we also compute partial sums of conditional expectations 
that will be required in the M-step. 
These are denoted by $s_{1il}^{(k)}=\sum_{j\in J_l} \tau_{ij}^{(k)}$,
 $\bs_{2il}^{(k)}$, etc. The expression for these partial sums 
depend on the component density. In the case of the FM-CFUST model, 
the eight partial sums are given by equations (27) to (34) in \cite{llm19}.

%\begin{comment}

\begin{algorithm}
\caption{Parallel E-step of multi-node EM algorithm}\label{A2}
\begin{algorithmic}[1]
  \For {thread $l \in \{1, 2, \ldots, m\}$} concurrently
		\For {$j \in J_1$}
			\If {k>1} {for $i \in \{1, 2, \ldots, g\}$}
				\State Compute $f_{ij}^{(k-1)}$.
				\State Calculate $f_j^{(k-1)} = \sum_{i=1}^n \pi_i^{(k-1)} f_{ij}^{(k-1)}$.
				\State Calculate $\tau_{ij}^{(k)} = f_{ij}^{(k-1)} / f_j^{(k-1)}$.
				\State Calculate $\ell_l^{(k-1)} = \sum_{j=1}^n \log f_j^{(k-1)}$. 
			\EndIf
			\State Compute conditional expectations $\be_{1ij}^{(k)}, \be_{2ij}^{(k)}, \ldots$.
			\State Compute partial sums $s_{1il}^{(k)}, \bs_{2il}^{(k)}, \ldots$ 
			\State Return conditional expectations and partial sums to the master thread. 
		\EndFor
  \EndFor
\end{algorithmic}
\end{algorithm}  

%\end{comment}

%--------------------------------%
\subsection{Parallel M-step}

The M-step can be inherently separated into $g$ threads. 
It is also possible to split into a larger number of threads 
by separately the calculations of each component of $\bPsi_i^{(k)}$ 
into a number of threads. However, given that the M-step expressions 
are often relatively inexpensive to evaluate, it may be preferable 
for all components of $\bPsi_i^{(k)}$ to be computed 
by the same thread. While the $g$ threads are computing $\bPsi_i^{(k)}$ 
simultaneously, the master thread computes $\ell^{(k-1)}$ 
by calculating the summation of $\ell_l^{(k-1)}$ 
returned by the $l$ threads after the parallel E-step. 
Once the calculation of $\bPsi_i^{(k)}$ is completed 
by the $g$ parallel threads, these can be combined into $\bPsi^{(k)}$ 
by the master thread. A summary of the parallel M-step 
of the multi-node EM algorithm is presented in Algorithm \ref{A3}.      

%\begin{comment}

\begin{algorithm}
\caption{Parallel M-step of multi-node EM algorithm}\label{A3}
Split the E-step results from $l$ threads by component
\begin{algorithmic}[1]
  \For {thread $i \in \{1, 2, \ldots, g\}$} concurrently
			\State Compute $\bPsi_i^{(k)}$ using $\be_{1ij}^{(k)}, \be_{2ij}^{(k)}, \ldots$.
			\State Return $\bPsi_i^{(k)}$ to master thread. 
  \EndFor
	\State Compute $\ell^{(k-1)} = \sum_{l=1}^m \ell_l^{(k-1)}$. 
	\State Combine $\bPsi_i^{(k)}$ into $\bPsi^{(k)}$. 
\end{algorithmic}
\end{algorithm}  

%\end{comment}

%--------------------------------%
\subsection{Stopping criterion}

We adopt the Aitken acceleration-based stopping criterion 
to determine whether the EM algorithm can be stopped 
after the $k$th iteration. The details are given in Algorithm \ref{A4}. 
It can be observed from Algorithm \ref{A4} that 
the calculations are rather simple 
and hence can be performed by a single thread 
(that is, the master thread).    

%\begin{comment}
\begin{algorithm}
\caption{Stopping criterion for multi-node EM algorithm}\label{A4}
\begin{algorithmic}[1]
	\State Calculate Aitken's acceleration 
		$a^{(k-1)} = \frac{\ell^{(k)}-\ell^{(k-1)}}{\ell^{(k-1)}-\ell^{(k-2)}}$. 
	\State Calculate $\ell_\infty^{(k)} = \ell^{(k-1)} + \frac{\ell^{(k)}-\ell^{(k)}}{1+a^{(k-1)}}$.
	\If {$|\ell_\infty^{(k)} - \ell^{(k)}|<\epsilon$} 
		\State Return $\bPsi^{(k)}$ and $\ell^{(k-1)}$.
		\State Terminate the algorithm.
	\EndIf
	\State Set $k$ to $k+1$. 
	\State Return to Algorithm \ref{A2}. 
\end{algorithmic}
\end{algorithm}

%\end{comment}

%---------------------------------------------------%
\section{Numerical experiments}

We performed numerical experiments to assess the performance 
of the multi-node EM algorithm under three different physical settings. 
The algorithm was implemented using both R and MATLAB 
and executed on machine(s) running Windows. 
The machine(s) have four physical cores. 
For the R implementation, the \texttt{parallel} package 
that is available in base R was used. 
For the MATLAB implementation, the Parallel Computing Toolbox was used. 
For a fair comparison between MATLAB and R, 
efforts have been made so that the R and MATLAB codes 
are almost direct transcriptions of each other. 
For example, the same set of commands were implemented 
for the computation of (multivariate) truncated moments. 
For illustration, the algorithm was implemented for the FM-CFUST model 
and applied to the Australian Institute of Sport (AIS) data \cite{B007}. 
The AIS data comprises $p=11$ body measurements on $n=202$ athletes. 
We apply the clustering algorithms to predict the gender 
of each athlete. For consistency, the same set of initial partitions 
was used and the number of components were fixed at $g=2$. 
The three settings for parallel computing are listed below.        
\begin{enumerate}
\item \emph{single-threaded}: using a single thread 
		on a CPU core on a single machine
\item \emph{multi-threaded}: using 2 to $12$ threaded 
		on a single machine. Note that the threads may be virtual threads.  
\item \emph{multi-node}: using multiple threads 
		from two or more machines connected through a local network. 
\end{enumerate}
For each of the above setting, the MATLAB and R implementations 
were ran separately on the same machine(s) for 100 replications each. 
We note that the multi-node version can be rather complicated 
to setup in R. As such, for the multi-node setting, 
we focus on the MATLAB implementation. 

As our main interest is the performance gain 
of the multi-node EM algorithm, details of the accuracy of 
the estimates and the clustering performance will not be reported here. 
However, we noted that the parameter estimates are almost identical 
to that obtained from the traditional (non-parallel) implementation, 
and all trials yielded the same final clustering.   

%--------------------------------%
\subsection{MATLAB versus R in single-threaded implementation}

The single-threaded version of the multi-node EM algorithm 
corresponds to the traditional implementation of the EM algorithm. 
This is the default setting in R and MATLAB. 
The mean and standard deviation (sd) of the run time for R 
are 2310.83 seconds and 62.04 seconds, respectively; 
see the third row of Table \ref{T1}. 
For MATLAB, the mean run time of the 100 trials are 2635.69 seconds, 
with sd being 142.63 seconds. 
On average, MATLAB appears to be slightly slower than R 
(approximately $14\%$) in this case. 
These results can be used as baseline measurements 
for comparison with multi-threaded and multi-node implementations.

%--------------------------------%
\subsection{MATLAB versus R in multi-threaded implementation}

\begin{figure}
	\centering
		\includegraphics[scale=.40]{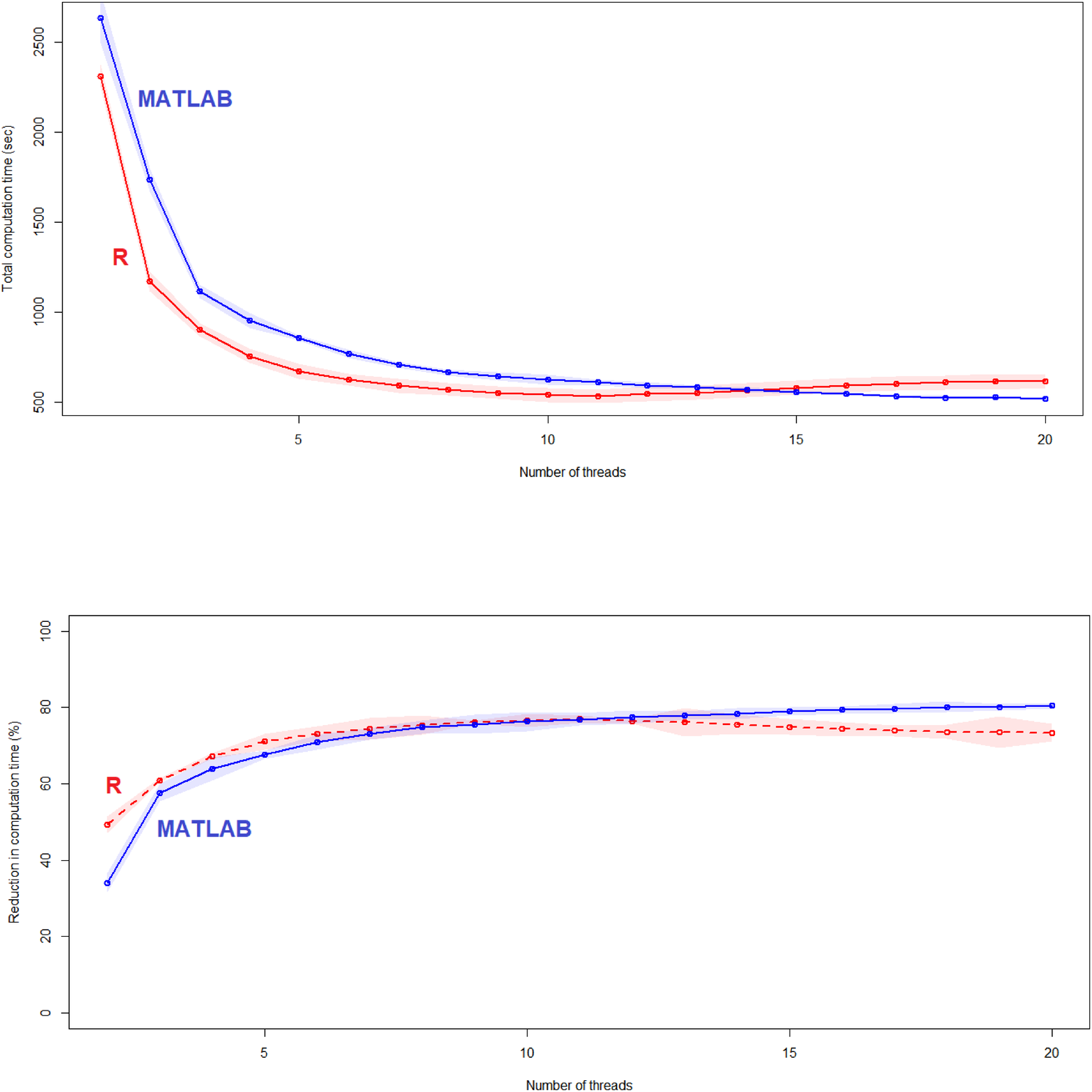}
	\caption{Performance gain of R and MATLAB implementations 
	of the multi-node EM algorithm on the AIS data set, 
	using $m=1, 2, \ldots, 20$ threads. The results for R is shown in red, 
	whereas the results for MATLAB is shown in blue. 
	Top panel: The circles represents the mean total computation time 
	(seconds) across 100 trials. Shaded region around the lines 
	indicate the standard deviation at each value of $m$. 
	Bottom panel: The percentage reduction in time (PRT) against 
	the baseline of $m=1$ were calculated for all cases of $m>2$. 
	The circles represents the mean PRT across 100 trials. 
	Shaded regions around the lines indicate the standard deviation of 
	PRT at each value of $m$. }
	\label{F1}
\end{figure}

With the multi-core/muti-threaded implementation, we would expect 
the computation to reduce as the number of threads increases. 
But linear reduction in $m$ should not expected as there are overheads 
associated with the setting up of the parallel process. 
In this experiment, we considered $m$ ranging from $2$ to $20$ threads. 
As there are only four cores in this machine, the threads are virtual 
when $m>4$. The total computation time (in seconds) in each trial 
and setting were recorded. For $m>2$, the reduction in time (in \%) 
against the baseline was computed using (total time in MATLAB - 
total time in R)/(total time in R) $\times$ 100. 
These results are reported in table \ref{T1} an displayed in 
Figure \ref{F1}. As can be observed the table and figure, 
significant reduction in time is achieved when the parallel 
implementation is used. With only two threads, 
the total time is reduced by $49\%$ and $34\%$ for R and MATLAB, 
respectively. At $m=4$ (the number of physical cores), 
a reduction of $67\%$and $64\%$ were achieved for R and MATLAB, 
respectively. However, the trend of decrease in total time 
starts to level out at around $m=10$ threads. 
For R, we could observe the total time even begins to increase mildly 
at $m=12$, possibly due to the overhead costs. On the other hand, 
the trend continued to decrease for large $m$ in the case of MATLAB. 
Although MATLAB implementation was slightly slower than the R 
implementation for small number of threads, 
it became faster than R for $m\geq 15$. 
 
In Figure \ref{F1}, the shaded region around each line 
is a representation of the standard deviation across the 100 trials 
in each setting. These trials indicate that 
the computation time for MATLAB seems to be more stable 
than R. This can also be gauged visually from the top panel of 
Figure \ref{F1}, where the R (red) line has a slightly broader 
shaded region than the MATLAB (blue) line, 
especially for larger values of $m$. 
The visual message from the bottom panel of Figure \ref{F1} 
suggests that both implementations have comparable 
percentage reduction in time, with MATLAB slightly more efficient 
when the number of threads is large.     

\begin{table}
	\centering
 \small\addtolength{\tabcolsep}{-5pt}
		\begin{tabular}{|c||c|c||c|c|}
			\hline
			\multirow{2}{*}{Number of threads ($m$)}	& 
			\multicolumn{2}{c||}{R} & 
			\multicolumn{2}{c|}{MATLAB} \\
			\cline{2-5}
			& Total time (sec) & Time reduction (\%) & Total time (sec) & Time reduction (\%) \\
			\hline\hline
	1 & 2310.82 & -- & 2635.69 & --  \\ 
  2 & 1171.65 & 49.30 & 1735.80 & 34.14 \\ 
  3 & 902.00 & 60.97 & 1115.17 & 57.69 \\ 
  4 & 756.48 & 67.26 & 952.46 & 63.86 \\ 
  5 & 669.72 & 71.02 & 854.23 & 67.59 \\ 
  6 & 623.03 & 73.04 & 768.07 & 70.86 \\ 
  7 & 591.08 & 74.42 & 706.83 & 73.18 \\ 
  8 & 570.85 & 75.30 & 664.17 & 74.80 \\ 
  9 & 552.47 & 76.09 & 643.54 & 75.58 \\ 
  10 & 539.92 & 76.64 & 625.11 & 76.28 \\ 
  11 & 532.29 & 76.97 & 608.86 & 76.90 \\ 
  12 & 544.34 & 76.44 & 594.21 & 77.46 \\ 
  13 & 551.34 & 76.14 & 581.34 & 77.94 \\ 
  14 & 565.21 & 75.54 & 568.77 & 78.42 \\ 
  15 & 580.05 & 74.90 & 556.78 & 78.88 \\ 
  16 & 593.98 & 74.30 & 545.38 & 79.31 \\ 
  17 & 603.16 & 73.90 & 534.20 & 79.73 \\ 
  18 & 609.62 & 73.62 & 523.43 & 80.14 \\ 
  19 & 613.66 & 73.44 & 525.70 & 80.05 \\ 
  20 & 615.02 & 73.39 & 516.34 & 80.41 \\ 
			\hline
		\end{tabular}
	\caption{Total computation time (sec) and time reduction (\%) 
	of the R and MATLAB implementation on a single multi-core machine. 
	The reported values are the mean and standard deviation (sd) 
	from 100 trials for each setting of $m=1, 2, \ldots, 20$. 
	The experiment applied the FM-CFUST model on the AIS data set.  }
	\label{T1}
\end{table}

%--------------------------------%
\subsection{MATLAB multi-node implementation}

For illustration, we tested the multi-node EM algorithm 
on a small local network of three machines with the same specifications. 
The number of threads is set to the total number of physical cores 
in the network. We recorded the total computation time of 100 trials 
in the case of one, two, and three machines. 
The results obtained by running on one machine is taken as 
the baseline for computing the percentage reduction in time 
for the case of two machines and three machines. 
With the multi-node setting, we would not expect a particularly good 
reduction in time for small data sets like in this experiment, 
as the high overhead associated with network communicating between 
the machines can overshadow the time gained by parallelizing 
the E- and M-steps. 
Indeed, we only observed a modest $2\%$ and $4.5\%$ 
mean reduction in time for the case of two and three machines, 
respectively. However, we expect these numbers to increase 
for larger data sets and its usefulness would be more apparent 
for models that are more computationally demanding.

%---------------------------------------------------%
\section{Conclusions}\label{sec5}

We have described a parallel implementation of the EM algorithm 
for the fitting of mixture models. The multi-node EM algorithm 
takes advantage of parallel computing to speed up 
the model fitting process. Quantitative comparisons were made 
between the MATLAB and R implementations of the same algorithm. 
We find that R is a little more efficient than MATLAB, 
despite the latter had built-in multi-threading capabilities. 
For both implementations (and in the single machine setting), 
a significant reduction of time was observed even for 
small number of parallel threads. For example, 
the total computation time for R had reduced to almost half 
when only two threads were used. 
For big data and/or models that are more computationally demanding, 
the multi-node setting could provide further reduction 
in computation time.

%---------------------------------------------------%

\end{document}